# Non-Einsteinian Viscosity Reduction in Boron Nitride Nanotube Nanofluids

André Guerra, Adam McElligott, Chong Yang Du, Milan Marić‡, Alejandro D. Rey\*, and Phillip Servio†

Department of Chemical Engineering, McGill University, Montréal, QC, Canada

\*alejandro.rey@mcgill.ca, †phillip.servio@mcgill.ca, ‡milan.maric@mcgill.ca

## Abstract

(1) Introduction: Nanoparticles have multiple applications, including drug delivery systems, biosensing, and carbon capture. Non-Einstein-like viscosity reduction has been reported in nanoparticle-polymer blends at low nanoparticle concentrations. More recently, a similar non-Einsteinian viscosity reduction effect has been observed in aqueous ultra-low concentration carbon-based nanofluids. (2) Methods: We use a boron nitride nanotube functionalized with hydrophilic groups in rheological experiments to investigate the viscosity reduction in ultra-low concentration nanofluids (0.1-10 ppm). We measure the dynamic viscosity in an air atmosphere and methane (0-5 MPag) at low temperatures (0-10 C).  (3) Results: A negligible effect on the temperature dependence of viscosity was found. Ultra-low concentrations of BNNT reduced the viscosity of the nanofluid by up to 29% at 10 ppm in the presence of methane. The results presented here were compared to similar studies on O-GNF and O-MWCNT nanofluids, which also reported significant viscosity reductions. (4) Conclusions: This work identified a non-Einsteinian viscosity reduction in BNNT nanofluids, which was exacerbated by methane dissolved in the nanofluid.

*Keywords*: non-Einstein, viscosity, reduction, nanoparticles, boron, nitride, nanotubes

## Graphical Abstract

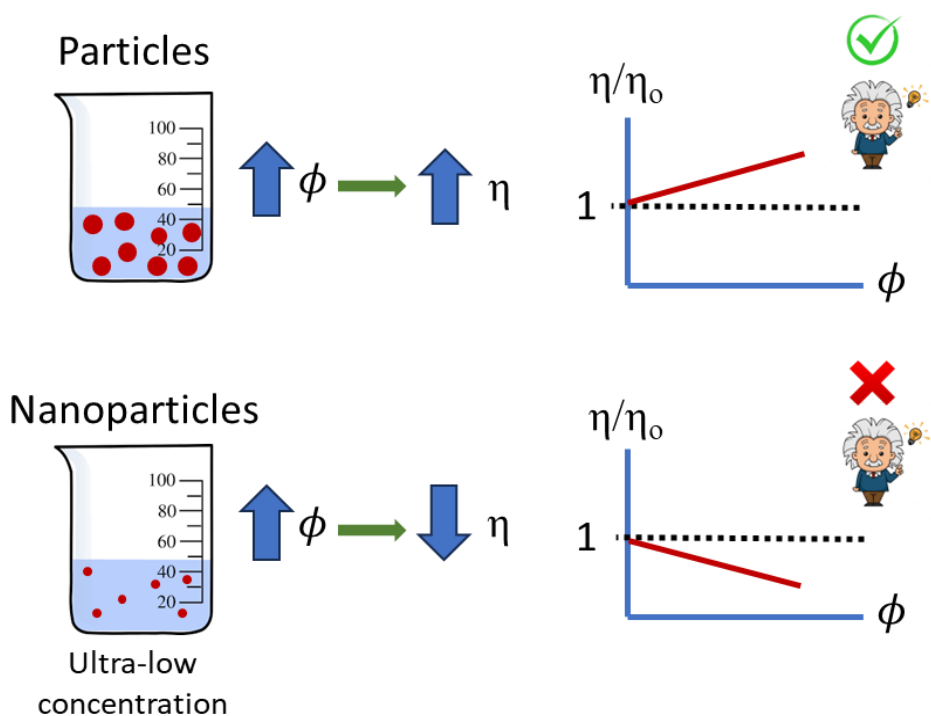

# 1. Introduction

Nanoparticles are a broad class of materials in which one of the dimensions is in the order of 100 nm or smaller[1]. Nanoparticles have several applications, from pharmaceuticals[2] to biosensing[3] and other process applications like carbon capture[4]. Pure carbon-based nanoparticles like graphene nanoflakes (GNF) and multi-walled carbon nanotubes (MWCNT) are hydrophobic. Their surfaces may be functionalized with oxygen or nitrogen groups to introduce hydrophilic properties, which improve water solubility and dispersion. More recently, oxygen and nitrogen functionalized GNF[5] and MWCNT[6] have been explored for their methane hydrate promotion abilities, which is observed at ultra-low concentrations (0.1-10 ppm by weight).

The conservation of energy equation, considering mechanical effects, includes the mechanical work term and its viscous contribution to the energy balance. This contribution represents the conversion of kinetic energy to heat through internal friction in the locally sheared fluid. This is an irreversible process, called viscous dissipation, and its magnitude is proportional to the fluid's viscosity. The internal friction arises from molecular interactions in the sheared fluid. One would expect that the addition of particles to a fluid would increase the internal friction, leading to greater viscous dissipation, and thus result in a greater effective viscosity in the mixture. This was one of the main propositions by Albert Einstein in his seminal work on Brownian motion. He proposed that the effective viscosity of a homogeneously dispersed solution of perfect spherical particles would be higher than the viscosity of the base fluid[7]. Moreover, in his work, he presented that the effective viscosity would be a function of only particle volume fraction and the viscosity of the base fluid. This idea has been further developed by Batchelor[8] and Krieger and Dougherty[9] to formulate models of the effective viscosity of nanofluids. However, all of these models are based on the original assumption that the addition of particles will inevitably increase the effective viscosity of the mixture, and thus fail to predict non-Einsteinian behavior - the decrease in effective viscosity as a result of the addition of particles to a fluid.

Previous work has reported non-Einstein-like viscosity reduction in nanoparticle-polymer blends. In 2000, Roberts et al. reported a decrease in viscosity in a blend of poly(dimethylsiloxane) with trimethylsilyl-treated silicate nanoparticles[10]. In 2003, Mackay and coworkers used a suspension of polystyrene (PS) nanoparticles in linear polystyrene (PS) at low nanoparticle concentrations (0.005-0.1 wt./wt.) to study this effect. They reported a reduction in viscosity which was unexpected and disagreed with predictions from Batchelor and Einstein's theoretical models[11]. The work presented an increase in free volume in the nanopolymer blend to be the source of the viscosity reduction. In 2005, Tuteja et al. examined the same system as Mackay et al. (PS nanoparticles and linear PS blend) to investigate the role of entanglement and confinement on the viscosity of these blends at low nanoparticle concentrations[12]. This work extended the previous conclusions by Mackay et al. about the role of free volume in the non-Einstein viscosity reduction and identified that molecular confinement was required to observe this effect. Tuteja et al. concluded that the confinement of entangled polymers contributed to the viscosity reduction effect as it was not measured for lower polymer concentrations (no confinement), in fact, viscosity increased in this case[12]. In 2012, Kim et al. further explored the non-Einstein-like viscosity behavior using several highly entangled polymers and silica nanoparticles. This work suggested the nanoparticle effect on the nanopolymer blend to be effectively that of a plasticizer to the entangled polymer melt[13].

The systems investigated in the studies mentioned above were nanopolymer blends and indicated nanoscale effects on the macroscopic viscosity of the blend as generally dependent on the physical conditions of, and interactions between, the polymer and nanoparticle molecules. Essentially, these studies identified a deviation from the continuum treatment of particles by Einstein and Batchelor, which assumed particles to be much larger than the molecular size of the suspension fluid. More recently, non-Einsteinian viscosity reduction was reported in ultra-low concentration aqueous carbon-based nanofluids of GNF[14] and MWCNT[15]. These systems are fundamentally different from the nanopolymer blends above and the mechanism leading to viscosity reduction may also be different. Generally, hydrophobic nanoparticles such as GNF and MWCNT are sparingly dispersible in water and require surfactants to avoid agglomeration in solution. Alternatively, the hydrophilic functionalization of nanoparticles results in nanofluid solutions that can remain dispersed for up to two years[16]. In the works by McElligott et al., the GNF and MWCNT were functionalized with hydrophilic carboxyl, hydroxyl, and ether oxide groups through a thermal plasma decomposition process. They report a viscosity reduction of upward of 29% upon the addition of oxygen-functionalized GNFs (O-GNF) at ultra-low concentrations (0.1-10 ppm)[14]. In the case of oxygen-functionalized MWCNTs (O-MWCNT), the viscosity reduction observed was upward of 19%[15]. These works proposed a hypothesis for the source of the observed effect that involves the disruption of hydrogen bond networks and enhancement of density fluctuations in the nanofluid leading to the reduced effective viscosity.

There have been several reports of viscosity reduction in propylene glycol-based nanofluids, which include nanoparticles of copper (II) oxide[17], iron (III) oxide [18], and manganese ferrite[19]. Manikandan et al. studied the thermophysical properties of nanofluid systems of nano-sand (72.6% $SiO_2$, 11.5% $Al_2O_3$, and 5.5% $Fe_2O_3$) in propylene glycol[20]. Their well-dispersed hydrophilic nano-sand nanofluids interfered with the hydrogen bond network in propylene glycol, causing up to a 46% reduction in viscosity at 2 vol.%. In 2021, Yadav et al. examined the rheology of nanofluids composed of $CeO_2$ and $Al_2O_3$ in ethylene glycol at low concentrations (0.05-1 vol.%)[21]. This work reported a non-Einsteinian reduction in viscosity for $CeO_2$ nanofluids at all concentrations, for the $Al_2O_3$ nanofluids at the lowest concentration (0.05 vol.%) below 70 C, and under a few conditions (below 40 C) for the $CeO_2$-$Al_2O_3$ hybrid nanofluid. This work cites the disturbance in hydrogen bond networks by the hydrophilic nanoparticles as the source of this effect[21]. Most recently, Arole et al. reported reduced friction from Stribeck curves in $Ti_3C_2T_z$ nanofluids in silicone and polyalphaolefin (PAO) oils. This resulted in a reduction in viscosity in the MXene nanofluid compared to its pure base-fluid by 12.3% in silicone oil (at 0.09 wt.%) and by 18.1% in PAO (at 0.04 wt.%)[22]. These interesting findings generate further questions on the mechanism and what other nanoparticles may exhibit this effect at low and ultra-low concentrations.

The functionalization of boron nitride nanotubes (BNNT) with hydrophilic groups was reported to occur during a novel purification process for as-produced BNNT[23]. This led to the production of boron-based nanoparticles with improved dispersion in water. This work examines the viscosity of ultra-low concentrations of aqueous BNNT nanofluids. We investigate the viscosity at low temperatures (0-10 C) in an air and a methane atmosphere (0-5 MPag). In this work, we report a negligible effect of BNNT addition on the temperature dependence of the nanofluid compared to the base fluid, though a reduction of viscosity with increased BNNT loading is identified. In addition, the dissolution of methane in the nanofluid is found to reduce its effective viscosity. Finally, this work compares its results from BNNT nanofluids to the results from our previous work on O-GNF and O-MWCNT nanofluids at similar conditions and concentrations.

# 2. Materials and Methods

## 2.1 Experimental equipment

The experiments conducted in this work were performed in the setup depicted in Figure 1. The main measurement device was an Anton Paar MCR 302 shear rheometer equipped with a gas-pressurized high-pressure cell. The measurements were collected from a double-gap geometry (double annulus) rotational shear device. The system was pressurized with air or methane (99.99% pure methane gas from MEGS) from a cylinder, and the high-pressure cell was maintained at the experimental temperature by a Julabo F-32 refrigerator using 50 vol.% aqueous ethylene glycol as the refrigerant fluid. The water used to make the BNNT nanofluids in this work had a 10 ppb organic content and was obtained by reverse osmosis using a 0.22 μm filter.

**Figure 1: Experimental equipment setup schematic.** From left to right: a gas cylinder, a Julabo F-32 refrigerator, and an Anton Paar MCR 302 gas-pressurized rheometer fitted with a high-pressure cell.

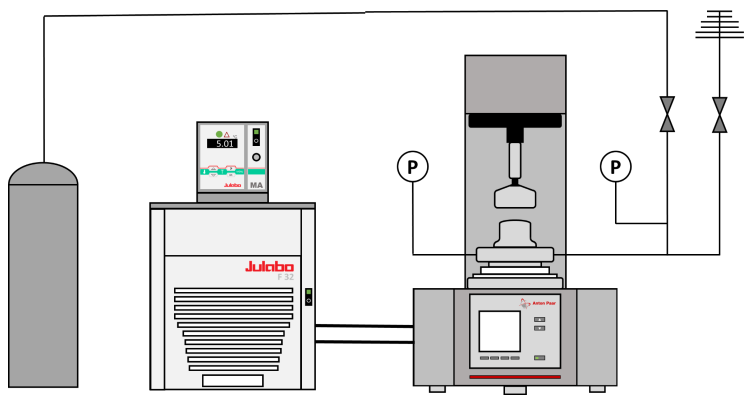

## 2.2. Production and characteristics of boron nitride nanotubes (BNNT)

The functionalized boron nitride nanotubes (BNNT) were previously produced and characterized by researchers at the National Research Council of Canada (NRC). A 50 ppm BNNT nanofluid sample was used in this study to produce the aqueous nanofluid solutions at several ultra-low concentrations through dilution (0.1, 0.5, 1, 5, and 10 ppm by weight). A detailed description of the production steps of the BNNTs used in this work is available elsewhere[23]. Here, we will provide only a summary of the process and the BNNT's properties. Hydrophobic as-produced BNNT containing approximately 50 wt.% impurities were purified in a three-stage process: (1) removal of hydrophobic impurities, (2) impurity size reduction, and (3) bromine treatment in water. In the first stage, hydrophobic impurities are removed through an iterative mechanical stirring, skimming, water extraction, and filtration (metal mesh 30-50 μm holes) process that is repeated until a fibrous suspension in clear water is obtained. In the second stage, ultrasonication is used to reduce the size of the remaining impurities and to separate them from the fibrous BNNT material obtained in the first stage of purification. Ultrasonication and vigorous mechanical stirring are simultaneously used in an extraction process, where the BNNT material settles to the bottom of the beaker while the impurities remain in the milky top layer, which is removed and decanted to recover the BNNT material. This process is repeated typically for 10 cycles. In the third stage, the purified BNNT material is treated with liquid bromine in an ultrasonication bath with constant magnetic stirring. The bromine reacts with the elemental boron producing boric acid and hydroboric acid. As a result, the final solution is near pH 1. The solid BNNT material is filtrated and collected. SEM analysis indicates an approximate 80 wt.% purity.

The purified BNNT material is soluble in water between pH 4 and 8, and it is readily precipitated out of solution upon pH reduction or increase outside this range. Guan et al. used a combination of thermal desorption in argon, thermogravimetric analysis (TGA), and Fourier Transform Infrared Spectroscopy (FTIR) to characterize the purified BNNT material[23]. It was found that BNNTs were functionalized with hydroxyl (OH) and amino ($NH_2$) groups. The addition of these functional groups on the surface of BNNTs was a result of the cleavage and hydrolysis of B-N bonds during the bromine treatment. The degree of functionalization was estimated from the TGA to be 2 OH groups and 1 $NH_2$ group per 53 BN pairs[23]. The presence of these hydrophilic functional groups on the surface of purified BNNTs allows their participation in hydrogen bond networks and results in increased solubility in water.

## 2.3. Experimental procedure

The rheometer's high-pressure cell sample well was loaded with 7.5 mL of BNNT nanofluid (water base fluid) at several concentrations (0.1, 0.5, 1, 5, and 10 ppm by weight). In experimental tests involving a methane atmosphere, the cell's headspace was charged with methane gas at 1 MPag to purge any air. The pressure was held for one minute before the cell was depressurized. This was repeated for 5 cycles to ensure all air was purged. The temperature was held constant (within 0.1 C of the setpoint) at the desired test condition (0, 2, 4, 6, 8, or 10 C) by the Julabo F-32 refrigerator. In experimental tests involving increased methane pressure, the high-pressure cell headspace was charged with the desired methane pressure controlled by the gas cylinder regulator (1, 2, 3, 4, 5

MPag). Once experimental conditions were achieved, the Anton Paar MCR 302 rheometer collected dynamic viscosity measurements using a constant 400 s$^{-1}$ shear rate, the manufacturer's recommended shear rate for the double gap geometry measurement of low viscosity fluids (e.g., water). Viscosity measurements for each condition were collected for a ten-minute period to ensure a stable reading (eliminate methane dissolution effect). Additionally, this set of data allowed for a 95% confidence interval to be calculated for the measured viscosity of each set of conditions considered. The Anton Paar software RheoCompass v.1.25 was used for data collection, and data analysis was conducted through in-house developed software in MATLAB®.

# 3. Results and Discussion

## 3.1 Temperature effect

In this section, we discuss the temperature effect on viscosity in water-based BNNT nanofluids compared to pure water exposed to air at atmospheric pressure. An increase in temperature is well known to decrease the viscosity of liquids. This is also true for nanofluids generally, and it is particularly the case for BNNT nanofluids at higher concentrations (0.001-0.03 wt.%)[24]. In liquids, increased temperature leads to higher molecular kinetic energy and increased Brownian motion leading to lower intermolecular interactions (including hydrogen bonding) and reduced resistance to shearing - lower viscosity. The viscosity measured in this work for pure water systems decreased linearly with temperature for all conditions, and this was also the case for all BNNT concentrations considered (Figure 2a). The addition of BNNT to water did not change the temperature dependence of viscosity observed in the baseline (pure water). The relative viscosity was calculated in this work as the ratio of the system's viscosity at each BNNT concentration to the viscosity of the baseline (pure water) for each temperature-pressure condition. The relative viscosity plot (Figure 2b) further demonstrates the absence of a considerable temperature effect at each BNNT concentration. The ratio used in the calculation of relative viscosity normalized the viscosity differences across temperatures for each concentration of BNNT. It is evident from Figure 2b that the data series for each BNNT concentration is relatively flat - indicating a negligible effect on viscosity arising from temperature change. This same observation was reported in our previous work examining O-GNF[14] and O-MWCNT[15]. More interestingly, however, Figure 2b demonstrates a non-Einsteinian viscosity reduction - the viscosity of water decreased with the addition of BNNT, and this effect was exacerbated with increased BNNT loading. This effect was not present for the lowest concentration (0.1 ppm) BNNT nanofluid, but all higher concentrations (0.5-10 ppm) exhibited a non-Einsteinian viscosity reduction. This effect will be further discussed in section 3.4.

**Figure 2: Viscosity of water-based BNNT nanofluid at several concentrations in air at atmospheric pressure.** Temperature and concentration effect of these systems on the (a) viscosity and (b) relative viscosity (normalized with 0 ppm). The error bars represent 95% confidence intervals and linear regressions are provided for each condition.

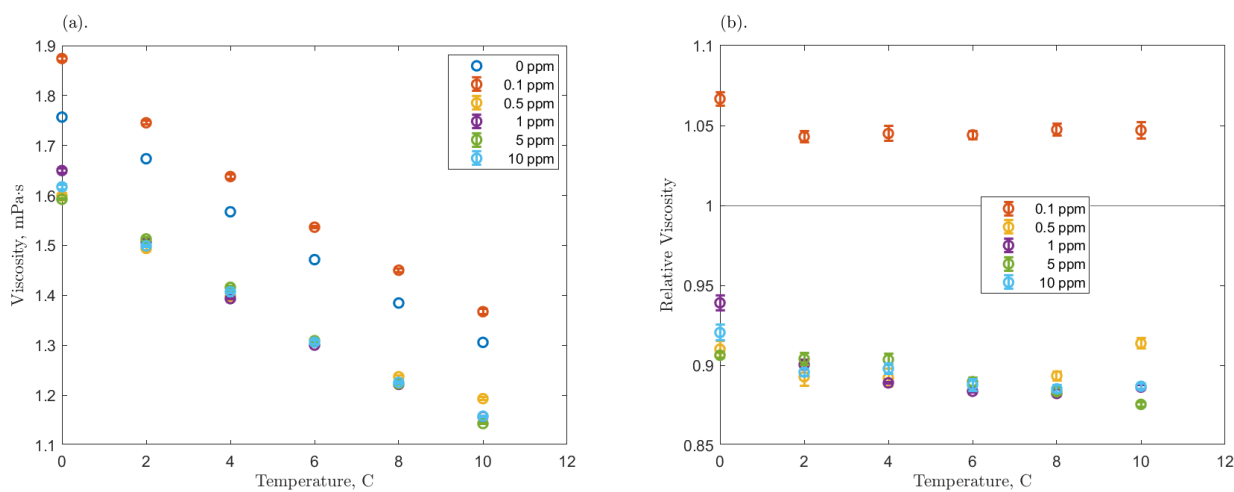

## 3.2. BNNT concentration effect

In this section, we discuss the concentration effect of BNNT in water under an air atmosphere at various temperatures. As briefly mentioned above, Figure 2b presents the relative viscosity of the BNNT nanofluids studied herein. All BNNT concentrations above 0.1 ppm exhibited a reduced viscosity when compared to the pure water system at equivalent conditions. The measured viscosity was approximately 11% lower than pure water on average across temperatures for 0.5, 1, 5, and 10 ppm systems. Conversely, the viscosity of the 0.1 ppm BNNT nanofluid was measured to be approximately 5% higher on average across temperatures (Figure 2b). Concentrations higher than 10 ppm may achieve further reductions, but there may be a threshold concentration above which this non-Einsteinian behavior may become dominated by larger-scale effects and the viscosity may increase as expected. However, in the data presented in this work, there may be an optimal cost-effect relationship at 0.5 ppm as it was capable of achieving approximately the same viscosity reduction as the 10 ppm system.

Comparatively, the viscosity reduction effect in O-GNF systems in an air atmosphere was lower than the BNNT systems studied in the present work. In our previous work, aqueous O-GNF systems (0.1-10 ppm) had an average viscosity reduction of approximately 3%, with the smaller reductions measured at 0.1 and 10 ppm, and the larger reductions at 1 and 5 ppm[14]. Moreover, the reduction in viscosity reported for O-MWCNT systems in an air atmosphere was in agreement with the effect observed here, but less pronounced than the results presented in Figure 2a. McElligott et al. reported that aqueous O-MWCNT solutions between 0.1 and 5 ppm achieved an average 4% reduction in viscosity across all temperatures (0-10 C), while 5 and 10 ppm systems were reported to result in approximately 5% higher viscosity than the baseline[15]. These results are similar to the effects measured in the BNNT nanofluids examined here. The differences in concentration effect may arise from the surface chemistry and geometric differences between the nanoparticles. The BNNT examined in the present work have an approximate 5.4 at.% surface functionalization of OH and $NH_2$ groups and on average they have a 5 nm diameter, 1-5μm length, and are made up of 2-5 walls[23]. The O-MWCNT referenced above, which is also a nanotube, has an approximate 21 at.% oxygen functionalization and an average 30 nm diameter and 10μm length[15]. Finally, the O-GNF has a fundamentally different geometry as it can be considered a mostly 2D nanoparticle - its planar dimensions are much larger than its thickness. It has an approximate 14.2 at.% oxygen functionalization, a planar dimension of 100 nm by 100 nm, and each nanoflake is composed of 5-20 atomic layers of graphene[25].

## 3.3. Methane atmosphere

Methane is a hydrophobic gas that may form hydrates in water under certain thermodynamic conditions and has been shown to have a weak effect on the viscosity of pure water at low temperatures under 5 MPag[26]. In this work, viscosity measurements were conducted at several methane pressures (0-5 MPag) for all BNNT concentrations considered (0-10 ppm). We present the 10 ppm trial in Figure 3 as it displayed the largest methane effect on the viscosity of the BNNT nanofluid system, but all others are included in the Supplementary Information (Figures A1-A3). In pure water, viscosity decreased consistently with temperature under isobaric methane conditions as expected (Figure A1a). Viscosity also decreased with pressure under isothermal conditions but with a weaker dependence than temperature. This is indicated by the higher temperature-viscosity (Figure A1a) and comparatively lower pressure-viscosity (Figure A1b) coefficients presented by the linear regressions of the viscosity data. The addition of BNNT was observed to cause a reduction in the measured viscosity of the nanofluid to a similar extent for 0.1 and 1 ppm (Figure A1 and Figure A2). However, upon the addition of BNNT, the temperature dependence of viscosity at both 0.1 ppm and 1 ppm was not impacted (Figure A2a and Figure A3a). At these concentrations, the viscosity decreased with temperature in agreement with the pure water case in a methane atmosphere (Figure A1a). This is consistent with previous reports for similar systems involving O-GNF[14] and O-MWCNT[15].

The effect of pressure on viscosity for 0.1 ppm and 1 ppm systems was also measured (Figure A2b and Figure A3b). The viscosity was reduced with increased pressure in the presence of BNNT in comparison to the pure water system (Figure A1b). Once again, the viscosity dependence on methane pressure was evidenced by the increasing viscosity-pressure coefficients across the progressively higher concentration of BNNT in the system. The effect of methane atmosphere pressure on the viscosity of 0.1 and 1 ppm BNNT nanofluids presented in this work was similarly reported for O-MWCNT aqueous nanofluid systems at the same concentrations[15]. This was not the case for another similar study performed with O-GNF. In the case of O-GNF, the effect of pressure on the viscosity of the nanofluids was reduced with increased concentration[14]. This will be further discussed in the next section which is focused on the non-Einsteinian viscosity reduction.

This work observed a negligible effect of the temperature dependence of viscosity at 10 ppm. This was similarly observed in previous work on examining O-GNF[14] and O-MWCNT[15] at the same conditions. In the present work, the largest effect of methane atmosphere pressure on the viscosity was captured for the 10 ppm system (Figure 3b). This is evidenced by the largest recorded pressure-viscosity coefficients from the linear regressions presented. It is worth noting that the effect of BNNT concentration on the temperature and pressure dependence of viscosity is coupled with the non-Einsteinian reduction in viscosity measured in this study and in our previous studies of O-GNF and O-MWCNT. Next, we discuss the non-Einsteinian viscosity reduction and the temperature and pressure dependencies of viscosity with calculated relative viscosity for average pressure and temperature, respectively, in an effort to clarify what is happening.

**Figure 3: Viscosity of water-based BNNT nanofluid at 10 ppm concentration in a methane atmosphere at several pressures.** (a) Temperature effect and (b) pressure effect. The error bars represent 95% confidence intervals and linear regressions are provided for each condition.

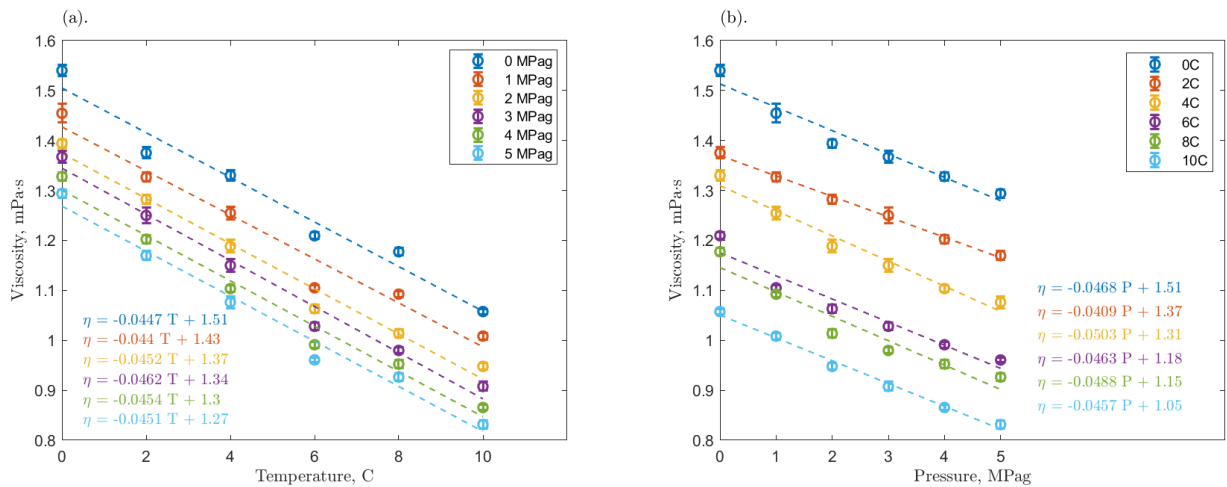

## 3.4. Non-Einsteinian viscosity reduction

In the beginning of the 20th century, Einstein introduced the idea of the Brownian motion of particles and the contribution of spherical particles to the viscosity of a fluid in which they are suspended[27]. The movement of particles in the fluid would cause them to experience a drag force as the fluid resisted their movement. This drag force is opposite to the fluid's internal viscous forces, which are proportional to its viscosity. Moreover, Einstein's model predicted an increase in internal viscous forces, and thus viscosity, as additional spherical particles were added to the suspending fluid. This theory relied on continuum assumptions including that the spherical particle's size was much larger than the molecular size of the fluid. In nanofluid applications, these assumptions may fail as the size of nanoparticles approaches the length-scale of molecular interactions. In these cases, the drag force experienced by the nanoparticle is not the dominant effect. In fact, for well-dispersed nanofluids, effects arising from intermolecular interactions, or their disruption, may be dominant at low concentrations.

As mentioned above, the present work identified a non-Einsteinian reduction in viscosity in the BNNT nanofluids considered. Firstly, it is evident from the results that a progressive reduction in nanofluid viscosity was measured as the loading concentration of BNNT increased in the nanofluid in the presence of methane (Figure 4). This effect was quantified through relative viscosity calculations as described in section 3.1 to decouple the temperature and pressure effects on the viscosity of the base fluid from the concentration effect. It is noteworthy that the 0.1 ppm nanofluid did not exhibit a non-Einsteinian reduction in viscosity in an air atmosphere (Figure 2b), but when the system is exposed to a methane environment a reduction in viscosity was measured (Figure 4). Methane is a hydrophobic gas that may be driven into the bulk of solution from the liquid-gas interface via hydrophobic interactions with nanoparticles - a shuttle effect[28]. Figure 4 demonstrates the combined effect of BNNT concentration and methane in reducing the viscosity of the nanofluid. The 10 ppm nanofluid exhibited the greatest reduction in viscosity of up to 29% (5 MPag). When compared to our previous works, O-MWCNT had a similar behavior to BNNT, but the viscosity reduction was not as large as reported in this work. The viscosity progressively reduced with higher O-MWCNT concentration (from 0.1 to 10 ppm) as observed in Figure 4, but the maximum reduction was reported to be only 19% at 10 ppm (6C)[15]. For the O-GNF work, the effect of nanoparticle concentration resulting in non-Einsteinian viscosity reduction was not the same. The maximum reduction was reported to be 29% at only 0.1 ppm (10C)[14]. Moreover, higher concentrations of O-GNF resulted in progressively higher viscosity. The non-Einsteinian viscosity reduction effect was still present, but the effect became weaker with increased concentration. These observations indicate that in the concentrations considered by our three recent works (0.1-10 ppm), the non-Einsteinian effect is stronger in the nanotubes (3D) when compared to the nanoflake (2D) - this may indicate the role of the geometry of the nanoparticle on this effect. Additionally, the higher probability for flow-induced alignment of nanotubes, compared to the nanoflake particles, may also contribute to sustained reduced viscosity before drag forces become dominant at higher concentrations leading to increased viscosity - returning to the Einsteinian regime.

In our previous work, we introduced a hypothesis for the source of the non-Einsteinian viscosity reduction[14]. We refer the reader to that work for a detailed description, here we provide a brief summary. Contrary to Einstein's model, at ultra-low concentrations, particles whose size is in a similar order of magnitude as the molecular size of the suspending liquid may not contribute sufficient internal friction to increase the effective viscosity of the bulk fluid. Hydrophobic interactions may disrupt hydrogen bond interactions in the base fluid, and the presence of nanoparticles may enhance density fluctuations, improving the dissipation of energy in the nanofluid and reducing its viscosity. The hypothesis is in agreement with previous research presented in section 1, involving oil-based nanofluids and other types of nanoparticles, suggesting the disruption of hydrogen bond networks in the base fluid to be a central factor giving rise to the viscosity reductions reported. We recommend computational work in the form of molecular dynamics simulations to examine the molecular interactions between the nanofluid components and the base fluid. This will help to improve our understanding of the effects reported in the present and previous works.

**Figure 4: The relative viscosity of water-based BNNT nanofluid at several concentrations.** Each value presented is calculated with the average viscosity at each: (a) pressure (0 to 5 MPag) and (b) temperature (0 to 10C).

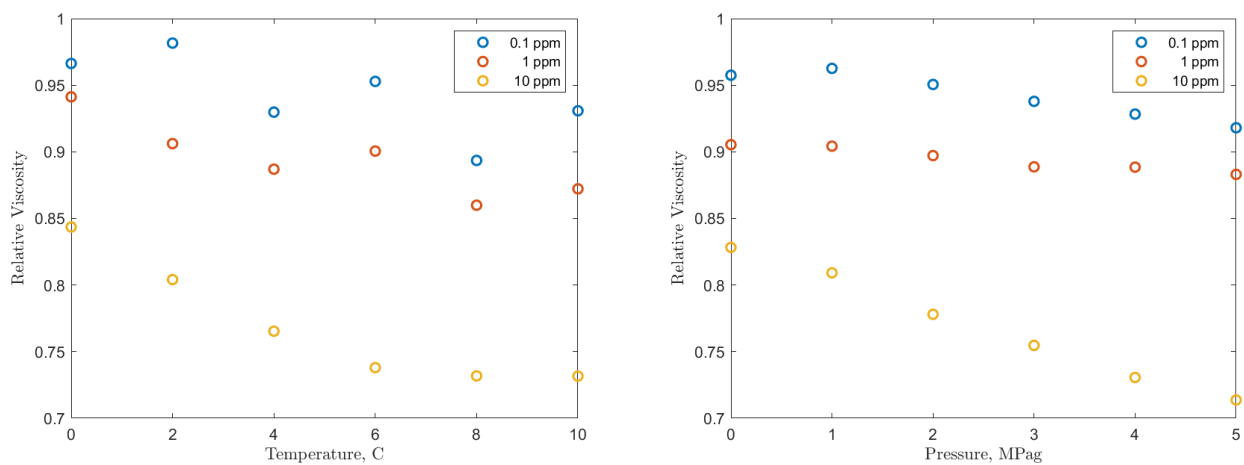

## 4. Conclusions and Future Work

This work examined the viscosity of ultra-low concentration aqueous BNNT nanofluids. The presence of BNNT in water at these concentrations had a negligible effect on the temperature dependence of the nanofluid's viscosity. This work measured a reduction in viscosity with increased loading of BNNT in the nanofluid. This non-Einsteinian viscosity behavior has not been reported for BNNT nanofluids, but it has been identified in water-based O-GNF and O-MWCNT nanofluids at similar concentrations. This effect was measured to be greater in BNNT nanofluids than in the O-GNF and O-MWCNT nanofluids. Considering these nanoparticles' different surface chemistry, this suggests that their geometric characteristics may also be contributing to this effect. The presence of a methane atmosphere was tested, and it was reported that the 10 ppm BNNT nanofluid demonstrated the highest effect on viscosity. This is likely due to the hydrophobic surface of the nanoparticles shuttling methane into the liquid phase, which introduces hydrophobic methane bubbles that further disrupt hydrogen bond networks. Throughout this work, a comparison between similar reports of non-Einsteinian viscosity reduction in O-GNF and O-MWCNT nanofluid systems was done. The differences in effect are likely due to varying hydrophobicity and other characteristics arising from the surface chemistry and geometry of the nanoparticles involved.

We suggest future work to explore the limits of the non-Einsteinian effect by considering higher BNNT concentrations. Moreover, nanoparticles with different geometries and amounts of functionalization should also be tested (e.g., fullerenes, graphene oxide, carbon dots). Future research should also consider the effect of these nanoparticles in ultra-low concentrations in oil-based nanofluids. Finally, computational studies (e.g., molecular dynamics) are required to study the molecular interactions between nanofluid nanoparticles and the base fluid to identify the molecular-scale sources of the macroscopic effect reported here and in our previous works - the non-Einsteinian viscosity reduction at ultra-low nanoparticle concentrations.

# Author Contributions

# Acknowledgments

The authors acknowledge the support from the Digital Research Alliance of Canada, Calcul Quebec, and WestGrid through computational resource grants, expertise, and technical support. The authors also thank the researchers from the National Research Council of Canada in the Securities and Disruptive Technologies Research Centre for providing the BNNT samples used in this study.

# Funding

Financial support for the work presented here was received from the Natural Sciences and Engineering Research Council of Canada (NSERC) through the Canada Graduate Scholarship Doctoral (CGS-D) award (A.G.), NSERC Discovery Grant number 206269 (P.S.), NSERC Discovery Grant number 206259 (M.M), NSERC Discovery Grant number 223086 (A.D.R.), Fonds de Recherche du Québec Nature et technologies (FRQNT) bourse de doctorat en recherche (CY.D.), from the McGill Engineering Doctoral Award (MEDA) (A.G. and CY.D.), and the James McGill Professorship (A.D.R.).

# Declaration of Competing Interest

The authors declare that they have no known competing financial interests or personal relationships that could have appeared to influence the work reported in this paper.

# Supplemental Information

Here, we provide all supplementary materials used in our analysis.

**Figure A1: Viscosity effect on pure water systems in methane atmosphere at several pressures.** (a) Temperature effect and (b) pressure effect. The error bars represent 95% confidence intervals and linear regressions are provided for each condition.

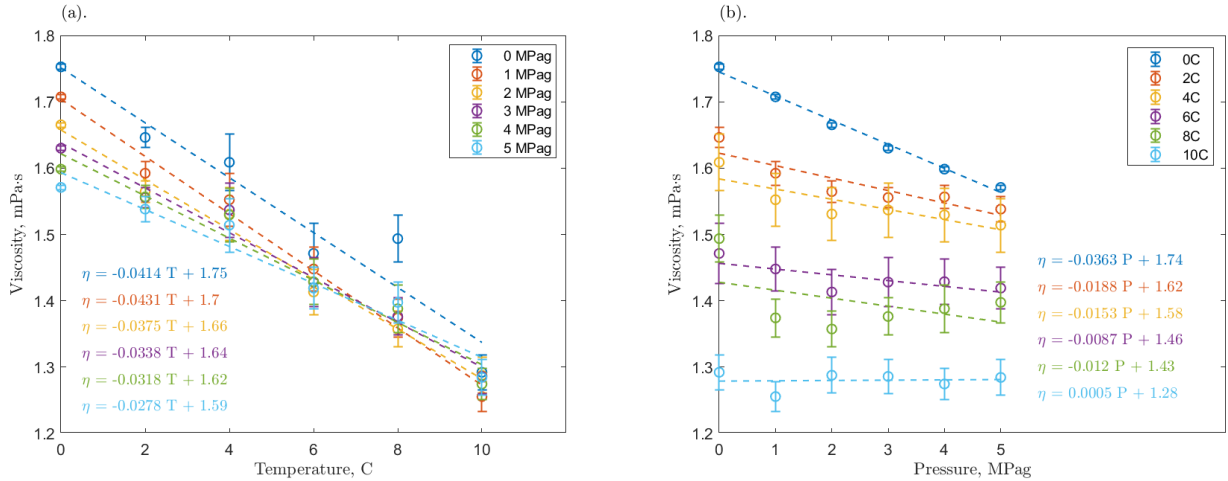

**Figure A2: Viscosity of water-based BNNT nanofluid at 0.1 ppm concentration in a methane atmosphere at several pressures.** (a) Temperature effect and (b) pressure effect. The error bars represent 95% confidence intervals and linear regressions are provided for each condition.

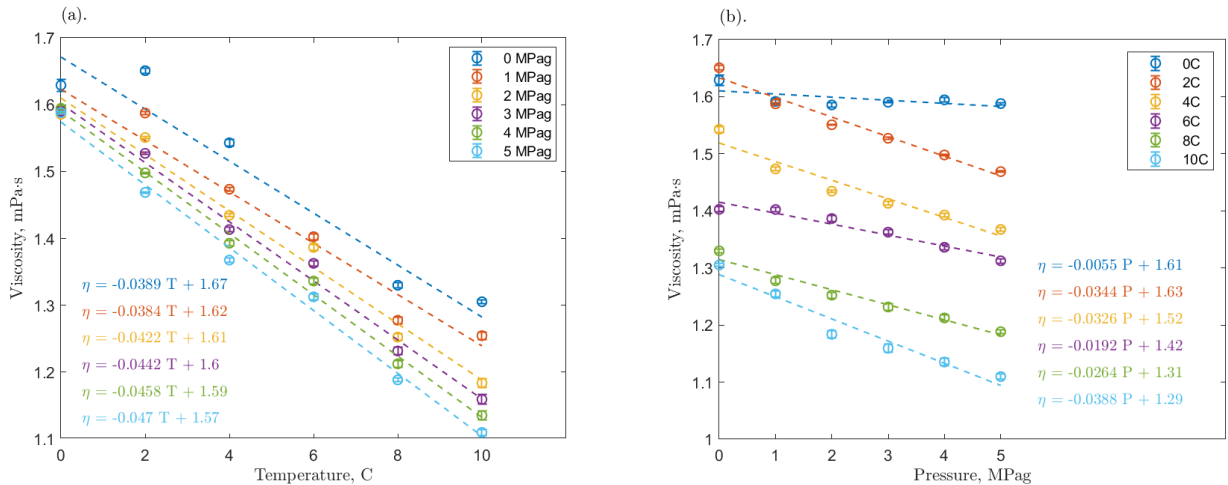

**Figure A3: Viscosity of water-based BNNT nanofluid at 1 ppm concentration in a methane atmosphere.** (a) Temperature effect and (b) pressure effect. The error bars represent 95% confidence intervals and linear regressions are provided for each condition.

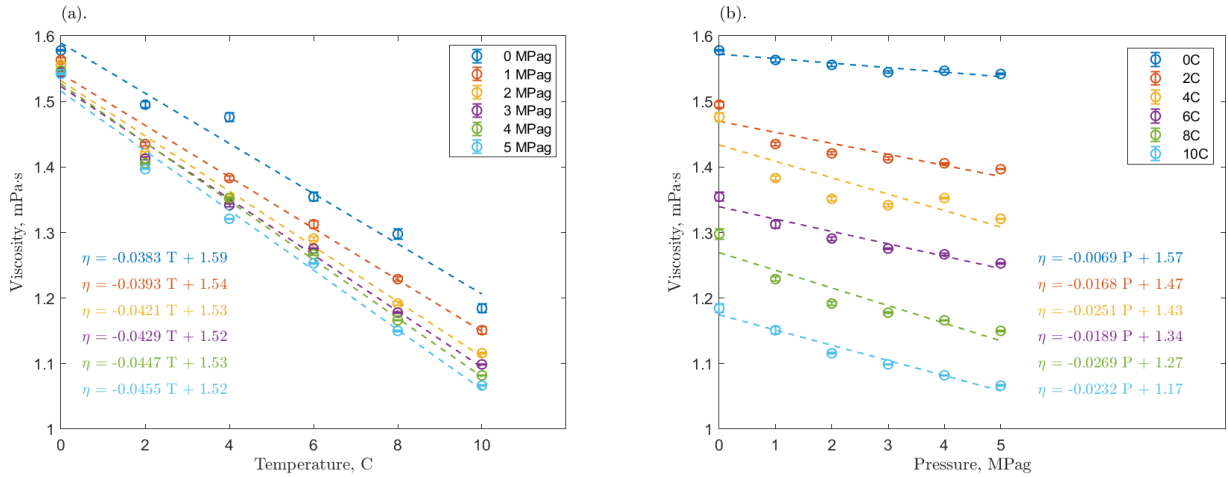